\documentclass{PoS}

\title{A suspected Dark Lens revealed with the e-EVN}

\ShortTitle{Dark Lens with the e-EVN}

\author{\speaker{Zsolt Paragi}\thanks{e-VLBI developments in Europe are supported by the EC DG-INFSO funded
        Communication Network Developments project 'EXPReS', Contract No. 02662.
        The European VLBI Network 
        is a joint facility of European, Chinese, South African and other radio 
        astronomy institutes funded by their national research councils.
        The Westerbork Synthesis Radio Telescope is operated by the ASTRON
        (Netherlands Institute for Radio Astronomy) with support from the
        Netherlands Foundation for Scientific Research (NWO). SF and ZP acknowledge partial support 
        from the Hungarian Scientific Research Fund (OTKA, grant no. K72515).
        }\\
        $^1$JIVE, Dwingeloo, Netherlands\\
        $^2$MTA Research Group for Physical Geodesy and Geodynamics, Penc, Hungary\\
        E-mail: \email{zparagi@jive.nl}}

\author{S\'andor Frey\\
        $^1$F\"OMI Satellite Geodetic Observatory\\
        $^2$MTA Research Group for Physical Geodesy and Geodynamics, Penc, Hungary\\
        E-mail: \email{frey@sgo.fomi.hu}}

\author{Bob Campbell\\
        JIVE, Dwingeloo, Netherlands\\
        E-mail: \email{campbell@jive.nl}}

\author{Attila Mo\'or\\
        MTA Konkoly Observatory, Budapest, Hungary\\
        E-mail: \email{moor@konkoly.hu}}

\abstract{The e-VLBI technique offers a unique opportunity for users
          to probe the milliarcsecond (mas) scale structure of unidentified 
          radio sources, and organise quick follow-up observations
          in case of detection. Here we report on e-EVN results for 
          a peculiar radio source that has been suggested to act as a  
          gravitational lens. However the lensing galaxy has not been 
          identified in the optical or the IR bands so far. Our goal
          was to look for an active galactic nucleus (AGN) in this 
          suspected dark lens system. The results indicate strong AGN 
          activity, and rule out the possibility that the radio source 
          itself is gravitationally lensed.\
          }

\FullConference{Science and Technology of Long Baseline Real-Time Interferometry: \\
The 8th International e-VLBI Workshop, EXPReS09\\
		 June 22 - 26 2009\\
		 Madrid, Spain}

\begin{document}

\section{A Dark Lens candidate}

In a campaign to study the optical properties of faint VLA FIRST 
(Faint Images of the Radio Sky at Twenty-Centimeters) sources with the 
Hubble Space Telescope, Russell et al. [\pos{1}] identified an optical arc
4 arcseconds away from FIRST J121839.7 +295325 (hereafter J1218+2953). 
There was no optical identification of the radio source itself.
The possible relation between the two objects was further investigated by
Ryan et al. [\pos{2}], who proposed that the arc may be a gravitationally lensed
image and the radio source may belong to the lensing object. There are two 
major difficulties with this interpretation. First of all the lensing galaxy
is not seen. Moreover the redshifts of the radio source and the optical arc
are not known. Ryan et al. [\pos{2}] estimated a redshift range of
$0.8<z<1.5$ for the radio source on the basis that the galaxy is not seen 
in the optical ($z>0.8$), but it has a relatively bright radio flux density ($z<1.5$). 
They carried out photometric redshift measurements of the optical arc with 
the SAO Multi-Mirror Telescope. A weak spectral break was seen at around
4300 \AA, which could be due to the Balmer/4000 \AA ~break at $z\sim0.13$, 
or due to the Lyman-break at $z\sim2.5$. Probability analysis showed 
that the Lyman-break is much more likely, which suggests that the arc is 
located at a high redhsift. Obviously, if the spectral break is related to 
the Balmer series then the redshift is much smaller and the two objects are 
unrelated.

Ryan et al. [\pos{2}] carried out gravitational lens modelling and found that
the mass distribution must be elliptical to produce such a long arc. However 
the observed sub-structure within the arc, and especially a bright knot at the
end, cannot be easily explained by gravitational lensing. 
They predict a secondary image which appears close to a pointlike source in the
HST image, but this is likely due to a "hot pixel". According to the model, the enclosed 
mass within the Einstein-radius of 1.3 arcseconds is $10^{12\pm0.5}M_{\odot}$.

It is not clear how such a massive object may remain hidden. Even if there is strong
obscuration by dust, there should be an infrared counterpart detected. IR imaging with 
the SAO Wide-field camera showed no galaxy with limiting magnitudes of $J=22.0$\,mag 
and $H=20.7$\,mag [\pos{2}]. Ryan et al. conclude that either there is an early-type
galaxy with significant amount of dark matter, or this could be a massive system
with an AGN that is completely obscured, with dynamic mass-to-light ratio exceeding
100 $M_{\odot}L_{\odot}^{-1}$.

\section{VLBI imaging of FIRST J1218+2953}

   \begin{figure*}
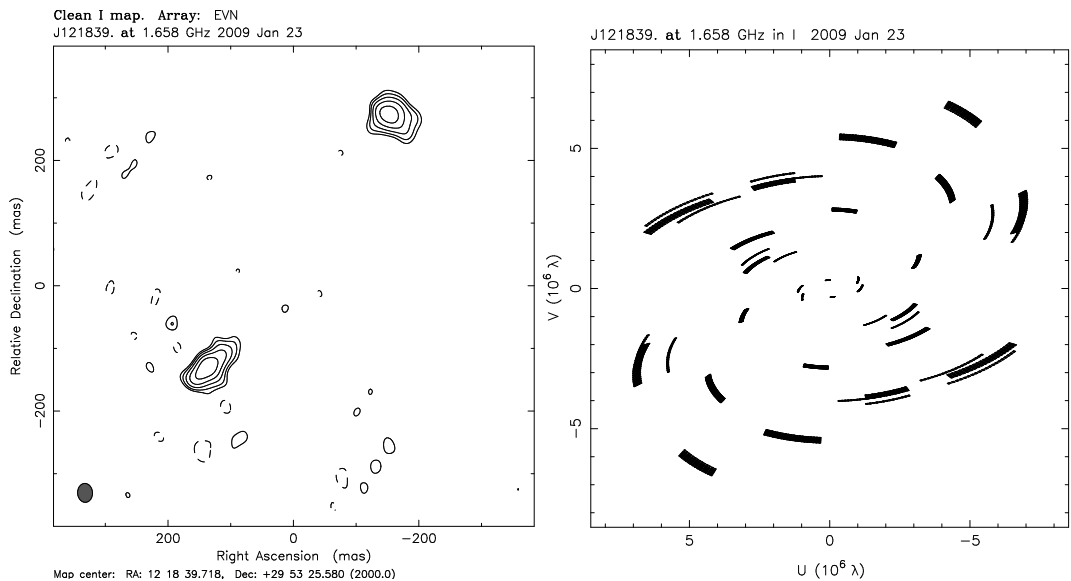

   \centering
   \vspace{20pt}
   \includegraphics[bb=33 120 582 714,clip,angle=0,width=7cm]{SHORT.ps}
   \includegraphics[bb=36 123 531 641,clip,angle=0,width=7cm]{UVPLOT.ps}
   \caption{\label{short} Left: naturally weighted image of J1218+2953 from 2 hours of short 
e-EVN observations at 1.6 GHz. The first contour is drawn at 0.4 mJy/beam, the following ones 
are multiples of this by $\sqrt{2}$. The beam was 30.3 $\times$ 23.8 mas, oriented at 
PA 3.3 degrees. The peak brightness was about 2.3 mJy/beam. The total cleaned flux density 
recovered was less then 15 mJy. Right: $uv$-coverage of the observations. The shortest MERLIN 
baselines (Jb2-Cm-Kn) are a great value for imaging of sources extending to several mas. 
However in this case the short observations were still not enough to recover most of the 
flux density.} 
    \end{figure*}

\subsection{Short e-EVN observations}

We applied for short e-EVN observations in December 2008, to test the AGN hypothesis
for J1218+2953. The total flux density of the source at 1.6 GHz is 33 mJy, which 
can be easily detected with limited EVN resources if it is compact. The observations
took place on 23 January 2009 at a data rate of 512 Mbps with 2-bit sampling, dual 
polarization, and lasted for 2 hours. The array consisted of Cambridge and  Knockin 
(MERLIN telescopes, limited to 128 Mbps), Jodrell Bank MkII, Medicina, Onsala, Torun and 
the Westerbork Synthesis Radio Telescope (WSRT). 
The target was phase-referenced to the nearby calibrator J1217+3007. 
The data were pipelined and then imaged in Difmap [\pos{3}]. The source was detected
and was resolved to two components, separated by about 500~mas (see Fig. \ref{short}). 

This result opened up new possibilities for the interpretation of the radio source.
Although it was not predicted from the "best-fit" gravitational lens model of Ryan et al.,
one may speculate that these two components are gravitationally lensed images of the
same background source that gives rise to the optical arc images, or perhaps a completely
unrelated background source. Alternatively, we may see a core-jet system or
a medium-size symmetric object (MSO). To distinguish between these scenarios we put in an
observing proposal by the 1 February 2009 deadline (just 8 days following the e-EVN
observations), for full-track e-EVN observations at 5 and 1.6 GHz.

\subsection{Follow-up experiments}

J1218+2953 was observed at 5 GHz on 24-25 March for 8 hours. The array this time
included the 100m Effelsberg telescope as well. Four telescopes (Ef, On, Tr, Wb) sent data 
to the correlator at 1024 Mbps rate, the rest at 512 Mbps or lower. The 1.6 GHz
observations were carried out on 21-22 April 2009 at 512 Mbps data rate. In the array
Knockin was replaced by Darnhall, the 76m Lovell Telescope was used instead of the MkII
in Jodrell Bank, and Arecibo joined as well (for 2 hours and 20 minutes). These observations 
and the data processing were similar to the short project described above. In addition, we 
reduced the synthesis array data from the WSRT that was obtained during the VLBI run.

The 5 GHz image resolves the South-East component into an elongated, slightly curved jet-like
structure, which does not point towards the North-West component (see Fig.~\ref{fulltracks}). 
There is no very compact component that could be firmly identified as a core. The total 
cleaned flux density is only 3 mJy compared to the total WSRT flux density of 9 mJy, indicating 
that most of the source is resolved out in this image. The 1.6 GHz image reveals an even more 
complex, but more continuous structure. The total cleaned flux density was about 20 mJy, close
to the total WSRT flux density of 27 mJy. 

\section{Interpretation of the results}

These preliminary e-EVN results show that the radio source near the optical arc has a complex
structure. The spectrum of the components is steep; that of the faint component near the phase 
centre is somewhat flatter. Because of the apparent quasi-continuous structure, the various 
components are likely not gravitationally lensed images of an unrelated background source. 

Comparing the total cleaned flux density of about 20 mJy to the WSRT flux density of 27 mJy, 
it is evident that most of the flux density is recovered at 1.6 GHz with the EVN. This indicates 
that the radio source and the optical arc cannot be lensed image pairs of the same background 
object, because in that case most of the radio flux density would be present near the optical 
arc since that image is strongly magnified (if the arc is indeed a lensed image). 

\begin{figure*}
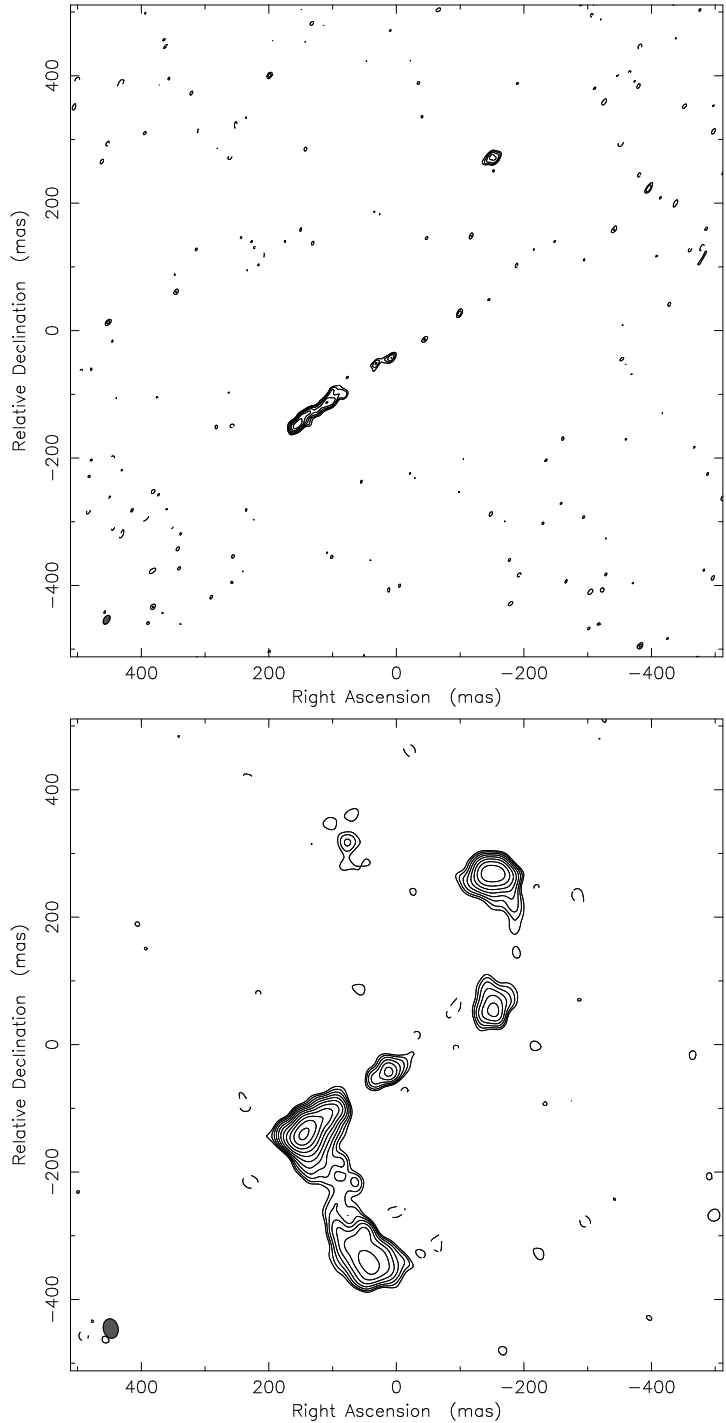

   \centering
   \vspace{20pt}
   \includegraphics[bb=33 138 582 685,clip,angle=0,width=9.5cm]{5GHz_map.ps}
   \includegraphics[bb=33 138 582 685,clip,angle=0,width=9.5cm]{1.6GHz_map.ps}
   \caption{\label{fulltracks} Full-track 5 GHz (top) and 1.6 GHz (bottom) e-EVN maps of J1218+2953.
   The peak brightnesses are 360 $\mu$Jy/beam and and 2.7 mJy/beam, respectively. The lowest contour
   levels are set at the 3~sigma noise level (49 $\mu$Jy/beam and 75 $\mu$Jy/beam, respectively)
   and they increase by a factor of $\sqrt{2}$. Both maps were naturally weighted; at 1.6 GHz
   a Gaussian taper was applied additionally, with half amplitude at 10 M$\lambda$. 
   The beam was 15.4 $\times$ 8.8 mas, oriented at PA $-$30.5 degrees at 5 GHz, and 
   30.9 $\times$ 23.3 mas, oriented at PA $-$13.3 degrees at 1.6 GHz.} 
    \end{figure*}

Further, gravitational lensing should be achromatic; thus were the pair of components to the 
SE and NW in the 1.6 GHz image (Fig.~\ref{fulltracks}, bottom panel) lensed images, the flux 
density ratio of the inner:outer components of each should be similar, a condition that is 
clearly violated. 

The most likely scenario is that the images show a single compact steep-spectrum (CSS) source 
(projected linear size < 20 $h^{-1}$ kpc), that might be categorized as a medium-size symmetric 
object (MSO, projected linear size > 1 $h^{-1}$ kpc) as well [\pos{4}]. There is thus evidence 
for AGN activity in the radio source. Note that if the optical arc is lensed by an object related 
to this AGN, then the mass-centre of the lens should be at the position of the AGN, which 
constrains further modelling of the system. 

Finally we note that this research benefited strongly from two aspects of the (e-)EVN: 
the additional short MERLIN spacings were of great use in recovering flux density on 
several-hundred mas scales, and the simultaneous recording of WSRT synthesis array data was 
very important for the intepretation of our results.   

\end{document}